\documentclass[prl,twocolumn,superscriptaddress,floatfix,10pt]{revtex4-1}
\usepackage{graphicx,bm,times}
\usepackage{amsmath}
\usepackage{amsfonts}
\usepackage{amssymb}

\usepackage{bm}
\usepackage{color}
\usepackage{times}

\usepackage{xcolor}
\usepackage[%
  colorlinks=true,
  urlcolor=blue,
  linkcolor=blue,
  citecolor=blue
  ]{hyperref}

\begin{document}

\title{Quantum oscillations probe the Fermi surface topology of the nodal-line semimetal CaAgAs}

\author{Y. H. Kwan}
\email[corresponding author:]{yves.kwan@physics.ox.ac.uk}
\affiliation{The Rudolf Peierls Centre for Theoretical Physics, Department of Physics,
	 University of Oxford, Oxford OX1 3NP, United Kingdom}

\author{P. Reiss}
\affiliation{Clarendon Laboratory, Department of Physics,
University of Oxford, Parks Road, Oxford OX1 3PU, UK}

\author{Y. Han}
\affiliation{Clarendon Laboratory, Department of Physics,
University of Oxford, Parks Road, Oxford OX1 3PU, UK}

\author{M. Bristow}
\affiliation{Clarendon Laboratory, Department of Physics,
University of Oxford, Parks Road, Oxford OX1 3PU, UK}

\author{D. Prabhakaran}
\affiliation{Clarendon Laboratory, Department of Physics,
University of Oxford, Parks Road, Oxford OX1 3PU, UK}

\author{D. Graf}
\affiliation{National High Magnetic Field Laboratory, Florida State University, Tallahassee, FL 32310, USA}

\author{A. McCollam}
\affiliation{High Field Magnet Laboratory (HFML-EMFL), Radboud University, 6525 ED Nijmegen, The Netherlands}

\author{S. A. Parameswaran}
\affiliation{The Rudolf Peierls Centre for Theoretical Physics, Department of Physics,
	University of Oxford, Oxford OX1 3NP, United Kingdom}

\author{A. I. Coldea}
\email[corresponding author:]{amalia.coldea@physics.ox.ac.uk}
\affiliation{Clarendon Laboratory, Department of Physics,
University of Oxford, Parks Road, Oxford OX1 3PU, UK}

\begin{abstract}
Nodal semimetals are a unique platform to explore topological signatures of the unusual band structure that
can manifest by accumulating a nontrivial phase in quantum oscillations.
Here we report a study of the de Haas-van Alphen oscillations of the candidate topological nodal line semimetal CaAgAs
using torque measurements in magnetic fields up to 45~T.
Our results are compared with calculations for a toroidal Fermi surface originating from the nodal ring.
We find evidence of a nontrivial $\pi $ phase shift only in one of the oscillatory frequencies.
We interpret this as a Berry phase arising from the semi-classical electronic Landau orbit
which links with the nodal ring when the magnetic field lies in the mirror ($ab$) plane. 
Furthermore, additional Berry phase accumulates while rotating the magnetic field for the second orbit in the same orientation
which does not link with the nodal ring.
These effects are expected in CaAgAs due to the lack of inversion symmetry.
Our study experimentally demonstrates that CaAgAs is an ideal platform for exploring the physics of nodal line semimetals
and our approach can be extended to other materials in which trivial and nontrivial oscillations are present.

\end{abstract}
\date{\today}
\maketitle

\textit{Introduction.---}The theoretical and experimental study of topological materials has been one of the most active fields in condensed matter physics over the last decade. Topological nodal-line semimetals (TNLSMs)~\cite{Burkov2011,Fang2015,Fang2016,Chiu2014} are recent additions to the roster of topologically nontrivial systems. The band structure of these  materials is characterized by the degeneracy of conduction and valence bands along a one-dimensional  line in the 3D Brillouin zone. Unlike the
point degeneracies in Weyl semimetals
that are robust to generic perturbations,  the stability of nodal lines requires additional symmetries, such as
inversion and time-reversal (as in Cu\textsubscript{3}PdN ~\cite{Kim2015,Yu2015}), reflection (e.g., PbTaSe\textsubscript{2}~\cite{Bian2016a}), or non-symmorphic symmetries  (as in ZrSiS~\cite{Schoop2016,Neupane2016}). TNLSMs can host nearly flat drumhead surface states~\cite{Bian2016b,Chan2016},
and are predicted to exhibit an array of unconventional electromagnetic properties~\cite{Ramamurthy2017} and correlation effects~\cite{Huh2016}.
Although in principle  the stability of TNLSMs with spin-orbit coupling (SOC) against forming topological insulators/semimetals is a matter of detail
that depends on the protecting symmetries, in practice the nodal-line description remains a good approximation for small
SOC.

Recent first-principles calculations and theoretical analysis suggest that the non-centrosymmetric pnictides \mbox{CaAgX (X = As, P)} are potential TNLSMs~\cite{Yamakage2016}. An unusual feature of these compounds is that at the intrinsic Fermi energy, no other structures apart from the nodal ring are predicted to be present. This allows experiments to directly examine properties of the nodal line without contamination from trivial Fermi pockets, in contrast to other proposed candidates. Hence, CaAgX has been dubbed the `hydrogen atom' of TNLSM phenomenology~\cite{Wang2017,Emmanouilidou2017}.
Transport studies indicate that 
the Fermi energy of CaAgAs is sufficiently close to the nodal line that the dispersion is within the  linear  regime,
whereas its phosphide counterpart exhibits significant band curvature~\cite{Okamoto2016}.
Therefore, experimental effort so far (ARPES~\cite{Nayak2018,Takane2018,Wang2017} and transport~\cite{Nayak2018,Okamoto2016,Emmanouilidou2017}) has focused on CaAgAs, which has a toroidal Fermi surface
due to intrinsic hole-doping caused by Ag deficiencies~\cite{Emmanouilidou2017}.

In this paper, we report a quantum oscillation study of single crystals of CaAgAs observed via torque measurements in magnetic fields up to 45~T.
By mapping the orientation-dependence of de Haas-van Alphen oscillations,
we provide evidence of a small toroidal Fermi surface with light cyclotron effective masses. 
We use a circular torus model and band structure calculations to discuss the origin of different
extremal orbits and their expected oscillations.
Building on this, we perform a phase analysis and find evidence of a nontrivial Berry phase arising from semi-classical electron orbits which link with the nodal ring. Our results provide experimental and computational
evidence of nontrivial  quantum oscillations spectra in an ideal TNLSM candidate.

\begin{figure*}
	\includegraphics[trim={0cm 1.7cm 0cm 0cm}, width=0.70\linewidth,clip=true]{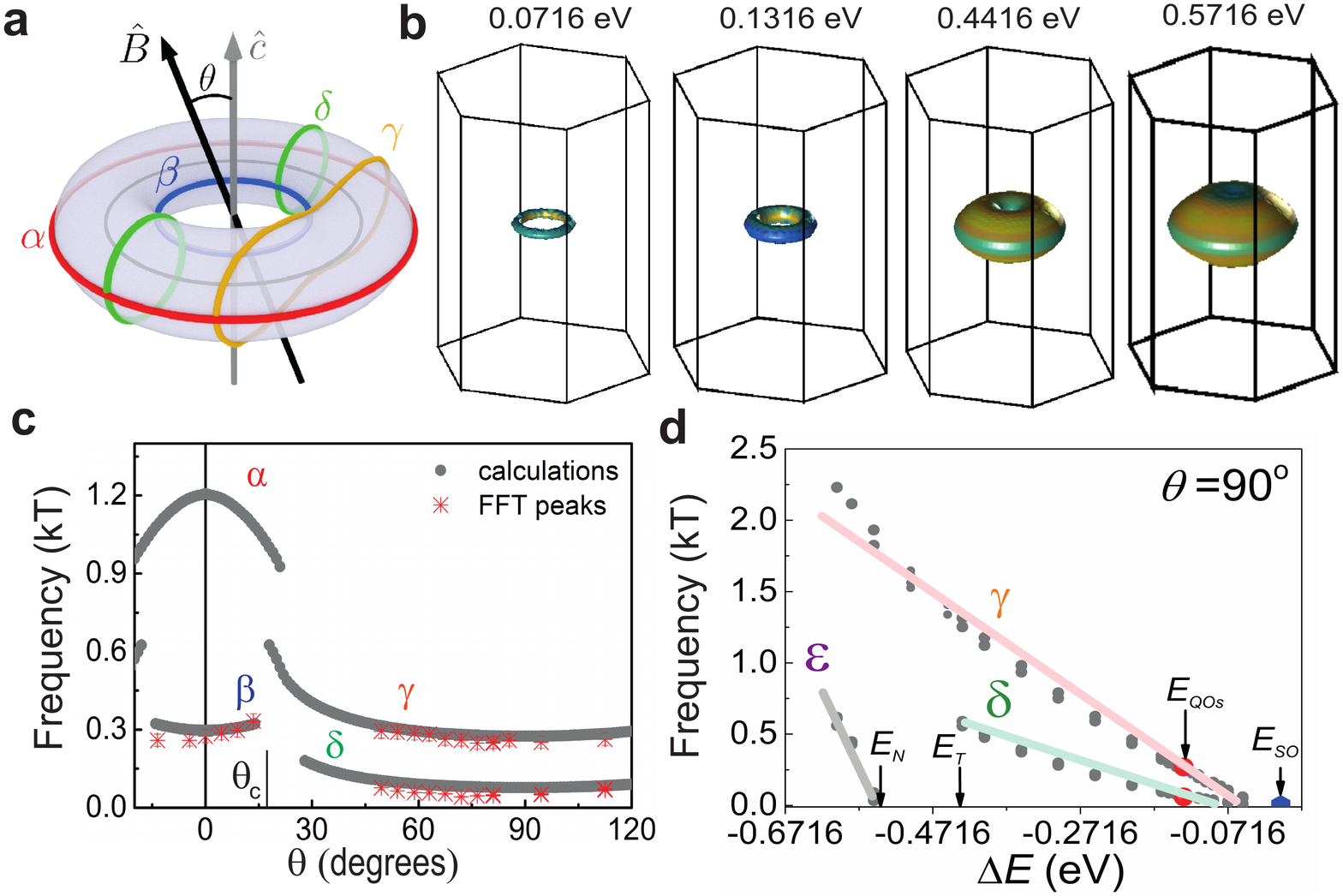}
	\caption{{\bf Fermi surface topology of CaAgAs.}
(a) Torus Fermi surface showing the relative orientation ($\theta$) of the magnetic field axis $\hat{B}$ with respect to the crystalline $c$ axis. $\alpha$ and $\beta$ are the semi-classical extremal orbits for $\hat{B}\parallel\hat{c}$ ($\theta=$~0$^\circ$), and $\delta$ and $\gamma$ for $\hat{B}\perp\hat{c}$ ($\theta=$~90$^\circ$).
Only $\delta$ links with the nodal ring and is expected to have a nontrivial Berry phase.
(b) Band structure calculation of the Fermi surface of CaAgAs for different values of hole-doping below the theoretical Fermi energy, plotted in the first Brillouin zone.
(c) Band structure calculations for the quantum oscillation frequencies as a function of $\theta$
with $E_{F}$ shifted
by $\sim$~0.13~eV due to hole-doping. Red asterisks are experimental data points.
(d) Predicted quantum oscillation frequencies as a function of hole-doping for $\theta=$~90$^\circ$. 
At perfect stoichiometry $E=E_{SO}$ there is no FS. 
$E_{QOs}$ is the Fermi energy based on experimental quantum oscillations. 
The torus closes at $E_T$, and a new band occurs inside at $E_N$. }
	\label{FigOrbits}
\end{figure*}

\textit{Fermi surface topology of CaAgAs.---}
CaAgAs 
crystallizes in the hexagonal space group P$\bar{6}$2m, and
the corresponding point symmetry group is D\textsubscript{3h}~\cite{Mewis1979}, which contains a mirror plane perpendicular to the $c$ axis but no inversion.
Without SOC, the mirror symmetry pins the nodal ring to the $k_z$~=~0 plane and protects it from gapping out since the crossing bands have opposite mirror eigenvalues.
In the presence of SOC, an energy gap of $\sim0.07$~eV opens and CaAgAs becomes a strong 3D topological insulator~\cite{Yamakage2016}.
However if the doping is greater than the SOC gap, many of the features of the underlying nodal-ring physics, such as the Fermi surface topology, remain relevant.

Quantum oscillations in 
TNLSMs have been predicted to exhibit a peculiar behavior \cite{Li2018,Oroszlany2018,Yang2018}
and can accumulate a topological Berry phase.
A qualitative understanding of the quantum oscillations of CaAgAs
can be gained by considering the limit of a circular torus FS [Fig.~\ref{FigOrbits}(a)], which is expected to capture the bulk behaviour of the nodal ring at small doping.
The fundamental frequency $F=\frac{\hbar}{2\pi e}A_{\it{k}}$ of quantum oscillations is determined by orbits on the
FS that enclose locally-extremal momentum-space area $A_{\it{k}}$~\cite{shoenberg_1984}.
Complex Fermi surfaces can have multiple
extremal orbits, and the experimental response will be a superposition of these oscillations.

Band structure calculations show the evolution
of the Fermi surface with hole-doping in Fig.~\ref{FigOrbits}(b).
The nodal ring of CaAgAs is centered at the Brillouin zone centre, $\Gamma$, and disperses linearly
\cite{Nayak2018,Takane2018,Wang2017}.
The toroidal FS increases in size with hole-doping, and eventually closes at an energy $E_T$.
Further hole-doping above 0.5~eV leads to the appearance of an additional band at $E_N$
inside the large Fermi pocket.

We find that both band structure calculations and the torus model (see Supplemental Material (SM) for details~\cite{SM})
predict two distinct oscillation frequencies for all magnetic field directions,
as shown in Fig.~\ref{FigOrbits}(c) (see also
Fig.~S3 and S5 in SM~\cite{SM}).
There is a critical angle, ${\theta_c}$,
that separates two different angular regimes.
Firstly,  at \emph{low angles} ($\theta < \theta_c$) close to  $\hat{B}\parallel\hat{c}$ ($\theta=$~0$^\circ$),
the extremal orbits are
$\alpha$ and $\beta$ shown in Fig.~\ref{FigOrbits}(a).
Neither of them link with the nodal ring.
With increasing angle $\theta$,
the larger orbit $\alpha$ monotonically decreases in frequency,
while the smaller orbit $\beta$ increases monotonically [Fig.~\ref{FigOrbits}(c)].
Secondly, for  \emph{high angles} ($\theta_c<\theta\leq$~90$^\circ$),
the handle orbit $\delta$ of the torus
is topological since it links with the nodal ring [Fig.~\ref{FigOrbits}(a)].
There is also an orbit $\gamma$ located away from the centre of the torus, which does not link with the nodal ring.
Both $\delta$ and $\gamma$ monotonically increase in frequency as $\theta$ is reduced from 90$^\circ$ [Fig.~\ref{FigOrbits}(c)].
In the vicinity of $\theta_c$, we expect magnetic breakdown~\cite{shoenberg_1984}
because different orbits approach each other closely in $k$ space.
This is expected to manifest itself in breakdown orbits with a quasirandom spectrum and a suppression of the oscillation amplitude~\cite{Alexandradinata2017}.

\begin{figure}[htbp]
	\centering
	\includegraphics[trim={0cm 1cm 0.7cm 0cm}, width=0.9\linewidth,clip=true]{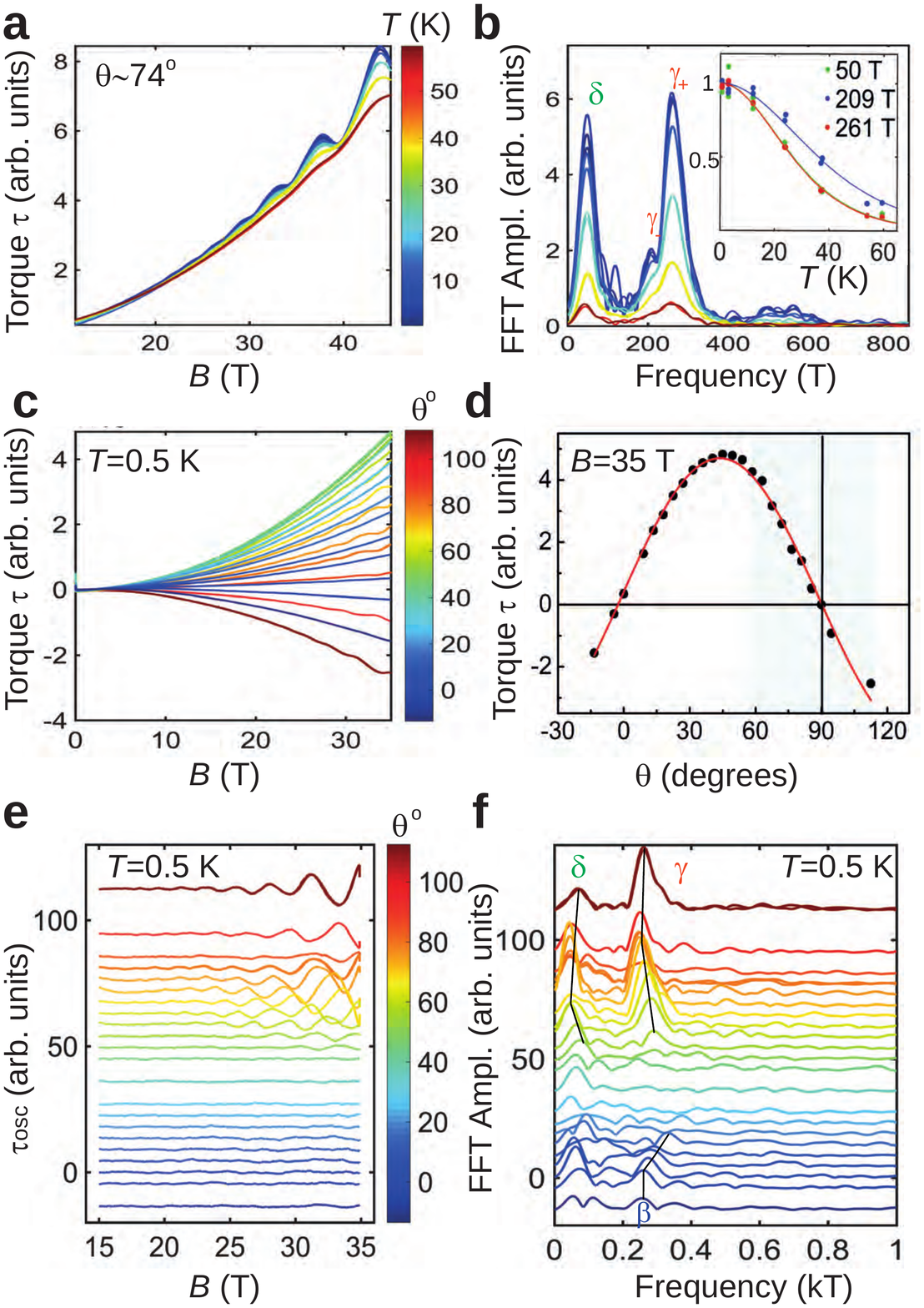}
	\caption{{\bf Quantum oscillations experiments in CaAgAs.}
(a) Total torque for a fixed field orientation ($\theta \sim$~74$^\circ$) 
at different constant temperatures.
(b) Fast Fourier transform frequency spectrum of (a) after subtraction of a polynomial background.
Inset shows the temperature dependence of different FFT amplitudes 
fitted to the damping factor $R_T$ to extract the effective masses.
(c) Torque data for different orientations in magnetic field at 0.5~K.
(d) Overall amplitude of torque in 35~T fitted by a $\sin2\theta$ dependence
indicates the symmetry axes where torque vanishes.
(e) Oscillatory torque, $\tau_{\rm{osc}}(B) $, obtained by subtracting a 
background from (c).
(f) Waterfall plot showing the evolution of the frequency spectrum as a function of $\theta$.
The angle is indicated by the vertical axis and the color bar in (e).}
	\label{FigWaterfall}
\end{figure}

\textit{Experimental details.---}
Single crystals of CaAgAs
were grown by the flux method \cite{Okamoto2016}
and form hexagonal rods along the $c$ axis.
Torque measurements were performed on a small single crystal of $<$~100~$\mu$m at NHMFL
in Tallahassee up to 45~T and HFML in Nijmegen up to 38~T using a piezocantilever, a single-axis rotator
(error less than 3$^{\circ}$), and a variable-temperature
cryostat. This crystal was verified by X-ray studies to be a single phase crystal (Fig.~S7).
Band structure calculations were performed using Wien2k with SOC
and GGA on a mesh containing $30 \times 30 \times 44$ $k$ points. Lattice parameters
are $a$~=~$b$~=~7.2040(6)~\AA~and $c$~=~4.2700(4)~\AA.

\textit{Quantum oscillations experiments in CaAgAs.---}
Figure~\ref{FigWaterfall} shows evidence of 
quantum oscillations in CaAgAs
from torque measurements as a function of magnetic
field for different orientations.
In order to isolate the oscillatory signal
, a quadratic polynomial is subtracted
from the raw data, since
the total torque is given by $\bm{\tau}=\mathbf{M}\times\mathbf{B}$.
The amplitude of quantum oscillations
can be described by the standard Lifshitz-Kosevich (LK) formula~\cite{shoenberg_1984},
as detailed in SM~\cite{SM}.
The oscillatory part of torque is given by the sum
of the contributions from all extremal orbits on the Fermi surface for a particular
orientation in magnetic field:
$\tau_{\rm osc}(B)
 \sim \sum_{i}A_i  B^{3/2}  R_T R_D R_S \sin \left(2\pi F_i/B+\phi_i  \right)$,
where $R_T$  accounts for finite temperature effects,
$ R_D$ for impurity scattering, and $R_S$ for spin-splitting effects.

Figure~\ref{FigWaterfall}(a) shows torque measurements up to 45~T
measured for different constant temperatures up to 60~K.
The FFT spectra indicate several frequencies
with a dominant frequency close to 260~T, a small shoulder around 210~T,
and a low frequency peak close to 50~T, as shown in Fig.~\ref{FigWaterfall}(b).
By fitting the $R_T$ damping term to the temperature dependence of the quantum oscillation amplitude, we extract very light masses of 0.12(1)~m$_e$ for
the 50~T and 260~T frequencies and 0.092(5)~m$_e$ for the 210~T frequency, as shown in the inset of Fig.~\ref{FigWaterfall}(b).
Similarly, fitting the $R_D$ term leads to a Dingle temperature of 65~K, a scattering time of 0.019~ps, and a mean free path of 420~\AA~ 
[see  Fig~S2(c)].
The proximity of the two largest frequencies
may indicate that they arise from Zeeman splitting of a single orbit. 
This scenario leads to a g-factor of $g\simeq$~7 for $\theta=$~74$^\circ$ (see SM~\cite{SM} for additional discussion on spin-splitting effects).

In order to understand the origin of the observed frequencies, we performed a series of measurements at various orientations $\theta$, as shown in Fig.~\ref{FigWaterfall}(c) [see also Fig.~S7(a)].
The angle-dependent Fourier spectra are summarized in Fig.~\ref{FigWaterfall}(f).
When the magnetic field is nearly perpendicular to the $c$ axis,
 there are two distinct oscillations at 50~T and 260~T, which we identify as $\delta$ and $\gamma$ respectively.
The position of $\theta=$~90$^\circ$ ($\hat{B}\perp\hat{c}$)  was deduced
based on the suppression of the torque at this high symmetry orientation, as shown in Fig.~\ref{FigWaterfall}(d).
 As $\hat{B}$ rotates from the $a$-$b$ plane to the $c$ axis, the $\delta$ and $\gamma$ oscillations increase
 in frequency as expected from band structure calculations in Fig.~\ref{FigOrbits}(c).
 The oscillations persist down to $\theta=$~50$^\circ$, 
 suggesting that the critical angle $\theta_c$ is below this value.
 At low angles close to $B||c$, we also observe an FFT peak around 260~T
 which is consistent with the $\beta$ orbit.

Comparing the experimental frequencies with
those predicted by band structure calculations shown in Fig.~\ref{FigOrbits}(d)
(additional simulations in Figs.~S3, S4, and S5 in SM~\cite{SM}),
we estimate that our system is hole-doped with an energy shift
 of $\Delta E \simeq$~0.1316~eV. This doping level generates
 frequencies of 67.5~T 
  for $\delta$
 and 245~T  
 for $\gamma$, in good agreement with the experimental data in Fig.~\ref{FigWaterfall}.
This energy shift predicts a critical angle $\theta_c \simeq$~21$^{\circ}$
and frequencies of $\sim$~1200~T for $\alpha$ and $\sim$~300~T for $\beta$, as shown
in Fig.~S3.
Based on these parameters,
we estimate the carrier density to be 5.2~$\times$~10$^{\text{19}}$~cm$^{-\text{3}}$,
close to the experimental value 7.5~$\times$~10$^{\text{19}}$~cm$^{-\text{3}}$ obtained via transport measurements~\cite{Okamoto2016}. 
ARPES measurements on CaAgAs have reported a wide range of hole-doping levels
with energy shifts from 0.05~eV~\cite{Takane2018} to 0.5~eV~\cite{Nayak2018,Wang2017}.

\begin{figure}[t]
	\includegraphics[trim={0cm 3.5cm 1.8cm 0cm}, width=0.9\linewidth,clip=true]{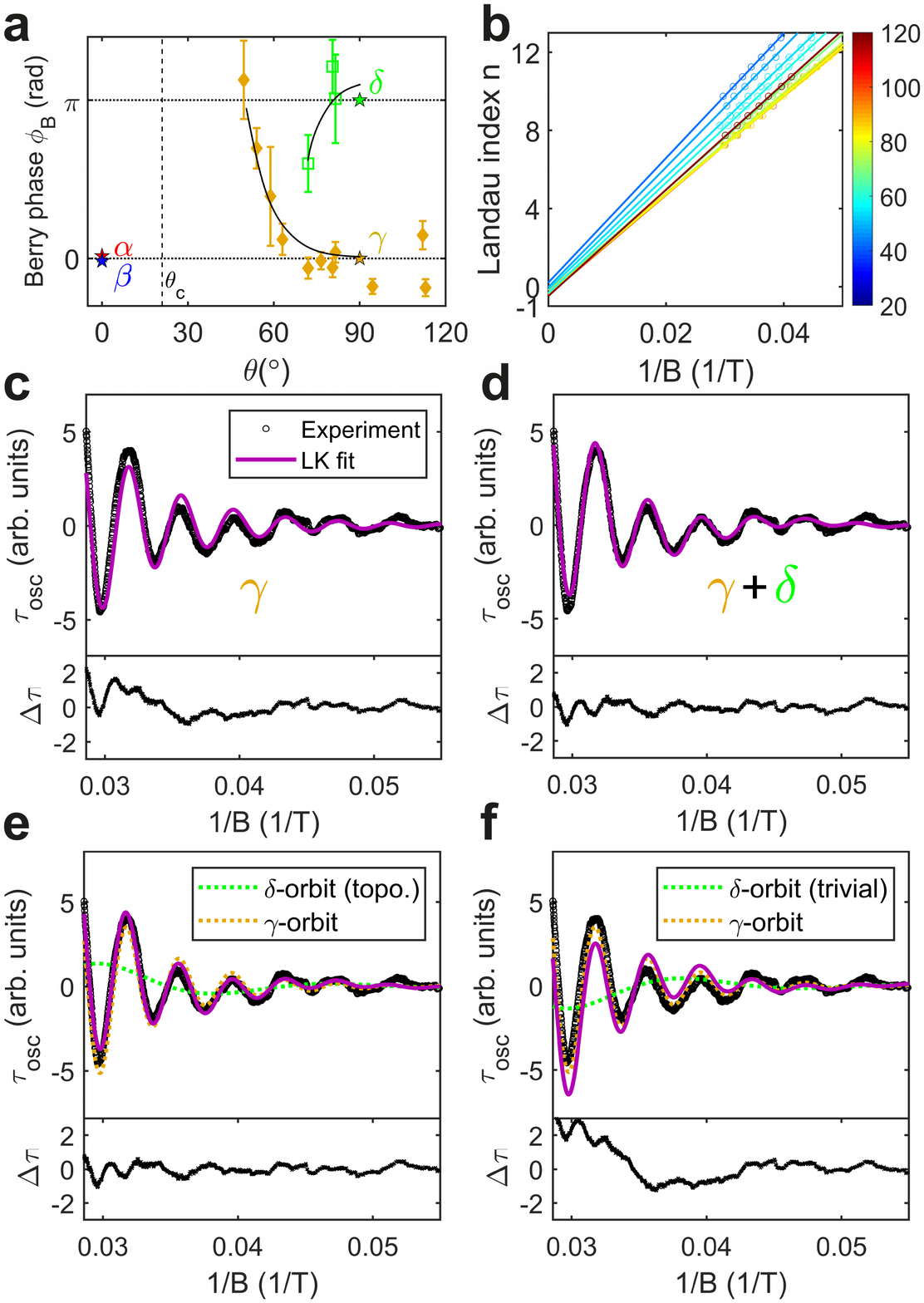}
	\caption{\textbf{Phase analysis of quantum oscillations.} (a) Berry phases extracted
		using a two-component LK formula. The stars represent theoretical predictions in the
		absence of SOC. (b) Landau fan diagram for $\gamma$ [see Fig~S6(c)]. (c) One-component LK fit to $\gamma$ for a field sweep at $\theta=$~81$^\circ$. 
Lower panel shows residuals. 
(d) Two-component LK fit including both $\gamma$ and $\delta$. 
(e), (f) shows the difference if $\delta$ has $\phi_B=\pi$
(topo.) versus $\phi_B=$~0 (trivial).
Dotted lines are the individual components of the LK fit.
The best fit is achieved in (e) indicating the topological nature of CaAgAs. }
	\label{FigPhases}
\end{figure}

\textit{Phase analysis of quantum oscillations.---}
The Berry phase, $\phi_B$, is a signature of nontrivial band topology,
and in TNLSMs with combined inversion and time-reversal symmetry $\mathcal{\hat{T}\hat{I}}$,
the Berry phase along any closed contour in momentum space is either $0$ or $\pi$.
Only contours that link nontrivially with the nodal ring pick up a $\pi$ Berry phase --
this is precisely the $\mathbb{Z}_2$ invariant describing TNSLMs protected by $\mathcal{\hat{T}\hat{I}}$.
On the other hand, CaAgAs is a non-centrosymmetric compound whose nodal ring (in the absence of SOC)
is protected instead by mirror symmetry, giving rise to a $\mathbb{Z}$ mirror invariant.
However,
the Berry phases remain quantized but {\it only} for those orbits defined at high symmetry directions, as 
shown in Table~I and derived in SM~\cite{SM}.

The phase factor of quantum oscillations for each
orbit $\phi_i=-\pi+\phi_B+\phi_{\text{3D}}$
depends on $\phi_B$. 
For hole carriers, $\phi_{\text{3D}}=\pm{\pi/4}$
depends on whether the extremal orbit has locally minimum or maximum area respectively~\cite{shoenberg_1984}.  
The phase shifts, in the absence SOC, can be predicted for each frequency branch \cite{Li2018}, and are summarized in Table~I.
\begin{table}
\caption{The various phases
 for the four types of extremal orbits of CaAgAs.
 The phase values listed in the last two columns are strictly 
 valid only for the magnetic field orientation shown in brackets.
 \label{TablePhases}}
	\begin{ruledtabular}
	\begin{tabular}{ccccc}
		Orbit & $\text{sgn}(\frac{dF}{d\theta})$ & $\phi_\text{3D}$ & $\phi_B$ & $\phi_i$  \\
		\hline
		$\alpha$ ($\theta=0^\circ$)     &$-$        & $\pi/4$            & $0$           & $5\pi/4$   \\
		$\beta$ ($\theta=0^\circ$)      &$+$        & $-\pi/4$           & $0$           & $3\pi/4$   \\
		$\gamma$ ($\theta=90^\circ$)    &$-$        & $\pi/4$            & $0$           & $5\pi/4$   \\
		$\delta$ ($\theta=90^\circ$)    &$-$        & $-\pi/4$           & $\pi$         & $-\pi/4$ \\
	\end{tabular}
	\end{ruledtabular}
	\end{table}
To address the topological character of $\delta$, we directly fit the oscillatory torque signal in the high-angle regime 
(over the field range 18-35~T as shown in Fig~\ref{FigWaterfall}(e) to minimize the spin-splitting effect)
with a simplified two-component LK formula.
This approach is preferred over a Landau fan analysis [Fig.~\ref{FigPhases}(b)] 
because of the multiple frequencies involved
(see SM~\cite{SM} for details of the fitting procedure).
Using non-linear least squares, we used the torque fitting formula
$\tau_{\rm osc}(B) $ to directly extract the phases of the two orbits close to $\theta=$~90$^\circ$, as
shown in Fig.~\ref{FigPhases}(a).
Torque is a signed quantity,
but in our measurements we are only certain of the magnitude of the signal.
Assuming that the sign of torque is negative (i.e.~CaAgAs is diamagnetic),
we find from direct fitting that the oscillatory torque data close to $\theta=$~90$^\circ$
is best described when the Berry phase of $\gamma$ is  $\phi_B=0$ and that of
$\delta$ is $\phi_B=\pi$, as shown in Fig.~\ref{FigPhases}(c-f). 
This finding provides strong evidence 
for nontrivial band topology in CaAgAs.
Furthermore, as the angle is reduced from $\theta=$~90$^\circ$, $\gamma$ accumulates additional Berry phase [Fig.~\ref{FigPhases}(a)]. This is consistent with the lack of inversion symmetry of CaAgAs, which allows Berry phases to vary away from high symmetry directions.

\textit{Conclusions.---} In this work,  we have used torque magnetometry to investigate the geometry and topology of the Fermi surface of the
proposed TNLSM CaAgAs. We analyze the experimentally observed quantum oscillations 
using theoretical input from both {\it ab initio} simulations and semiclassical analysis of a simplified model.  This allows us to link the
 observed oscillation frequencies  to the coexistence of multiple cyclotron orbits with both non-topological and topological Berry phases.
This provides strong evidence that the Fermi surface of hole-doped CaAgAs originates from an underlying topological line node at perfect stoichiometry (possibly weakly gapped by small SOC). We find no sign of any other bands near the Fermi energy, making CaAgAs an ideal setting to study TNLSMs and proximate insulating and semimetallic phases. Similar studies of samples 
with different doping levels may be able to reveal finer details of the nontrivial Fermi surface, 
and probe other fascinating phenomena enabled by the interplay of symmetry and band topology in CaAgAs and related materials.

\begin{acknowledgments}
	\textit{Acknowledgments.---}
	We thank Ni Ni for provision of related crystals that were not part of this study.
	This work was mainly supported by  EPSRC  (EP/I004475/1,  EP/I017836/1).
	A.I.C. acknowledges an EPSRC Career Acceleration Fellowship (EP/I004475/1). S.A.P. acknowledges support from the European  Research  Council  (ERC)  under  the  European Union Horizon 2020 Research and Innovation Programme  (Grant  Agreement  No.  804213-TMCS). Y.H.K. acknowledges support from NSF grant DMR-1455366 during the early stages of this project.
	A portion of this work was performed at the National High Magnetic Field Laboratory,
	which is supported by National Science Foundation Cooperative Agreement No. DMR-1157490 and the State of Florida.
	Part of this work was supported supported by
	HFML-RU/FOM and LNCMI-CNRS, members of the European Magnetic Field
	Laboratory (EMFL) and by EPSRC (UK) via its membership to the EMFL
	(grant no. EP/N01085X/1).
	We also acknowledge the Oxford Centre for Applied Superconductivity.
\end{acknowledgments}

\newcommand{\blue}{\textcolor{blue}}

\newcommand{\bdm}[1]{\mbox{\boldmath $#1$}}

\renewcommand{\thefigure}{S\arabic{figure}} 
\renewcommand{\thetable}{S\arabic{table}} 

\newlength{\figwidth}
\figwidth=0.48\textwidth

\setcounter{figure}{0}

\newcommand{\fig}[3]
{
	\begin{figure}[!tb]
		\vspace*{-0.1cm}
		\[
		\includegraphics[width=\figwidth]{#1}
		\]
		\vskip -0.2cm
		\caption{\label{#2}
			\small#3
		}
\end{figure}}

\newpage
\clearpage

\section{SUPPLEMENTAL MATERIAL}

\section{Theoretical Analysis of Berry Phases}
The Berry phase $\phi_B$ is often used as a signature of nontrivial band topology. In TNLSMs with combined inversion and time-reversal symmetry $\mathcal{\hat{T}\hat{I}}$, the Berry phase along any closed contour in momentum space is quantized to either $0$ or $\pi$. Only contours that link nontrivially with the nodal ring pick up a $\pi$ Berry phase -- this is precisely the $\mathbb{Z}_2$ invariant describing TNSLMs protected by $\mathcal{\hat{T}\hat{I}}$.

On the other hand, CaAgAs is a non-centrosymmetric compound whose nodal ring (in the absence of SOC) is protected instead by mirror symmetry, giving rise to a $\mathbb{Z}$ mirror invariant. However, we show below that when $\hat{B}$ is parallel/perpendicular to the $c$ axis, the Berry phases are still quantized. Namely for $\theta=90^\circ$, the $\delta$ orbit has $\phi_B=\pi$ while the $\gamma$ orbit has $\phi_B=0$. For $\theta=0^\circ$, the Berry phases are zero for both the $\alpha$ and $\beta$ orbits.

Consider a spinless non-degenerate band, which is valid in the limit of vanishing SOC. Under time-reversal symmetry, the band energy and Berry curvature satisfy
\begin{align}
E(\bm{k})&=E(-\bm{k}) \label{EqnTRSEnergy}\\
\bm{\Omega}(\bm{k})&=-\bm{\Omega}(\bm{-k}) \label{EqnTRSOmega}.
\end{align}
The other relevant symmetry operator is the reflection plane perpendicular to the $c$ axis, which enforces the following constraints
\begin{align}
E(\bm{k}_\parallel,k_z)&=E(\bm{k}_\parallel,-k_z) \label{EqnMirrorEnergy}\\
\bm{\Omega}_\parallel(\bm{k}_\parallel,k_z)&=-\bm{\Omega}_\parallel(\bm{k}_\parallel,-k_z) \label{EqnMirrorOmegaParallel}\\
\Omega_z(\bm{k}_\parallel,k_z)&=\Omega_z(\bm{k}_\parallel,-k_z) \label{EqnMirrorOmegaPerpendicular}
\end{align}
where $\bm{k}_\parallel=(k_x,k_y)$. The conditions on the energy are important in constraining the shape of semi-classical orbits which are always constant energy paths. The Berry phase for a closed contour $C$ in the Brillouin zone is given by
\begin{equation}
\phi_B=\int_{S} d\bm{k}\cdot\bm{\Omega}(\bm{k})\mod\,2\pi
\end{equation}
where $S$ is a surface with $C$ as its boundary. The above expression holds if the bounding surface does not encounter any gap-closing points.

\begin{figure}[htbp]
	\centering
	\includegraphics[trim={0cm 12cm 0cm 0cm}, width=1\linewidth,clip=true]{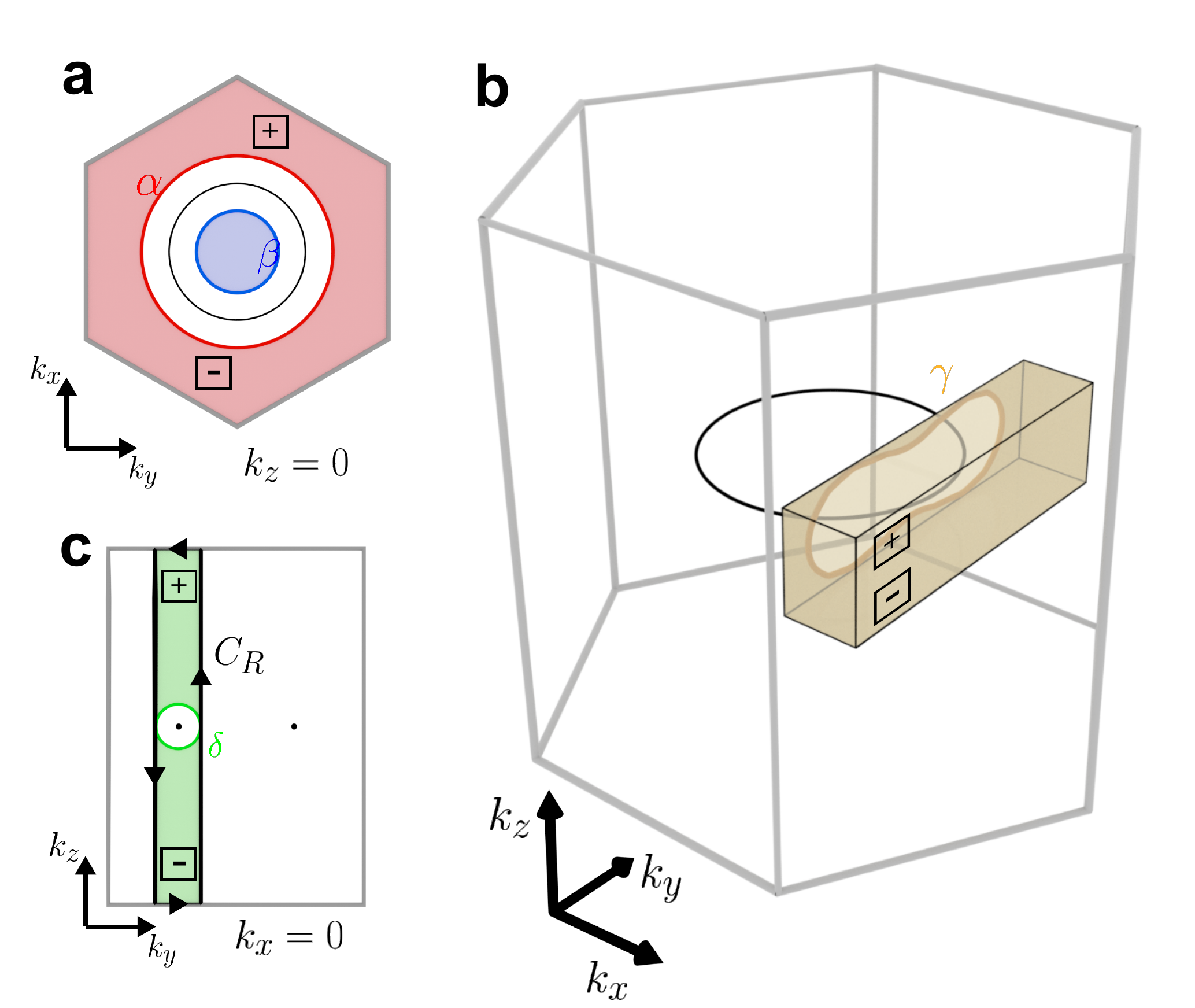}
	\caption{\textbf{Berry phase constraints}. (a) 2D slice of the Brillouin zone. For $\theta=0^\circ$, the integrated flux of Berry curvature through the shaded regions vanish. The $+$ and $-$ boxes show how the contributions cancel pairwise because of Equation \ref{EqnTRSOmega}. (b) For $\theta=90^\circ$, the Berry phase of the $\gamma$ orbit is also zero. (c) The contour $C_R$ has $\phi_B=\pi$ due to the band topology. This can be deformed into the $\delta$ orbit without net additional flux of Berry curvature. The two dots are gapless points arising from the nodal ring.}
	\label{FigSM:berry_curvature}
\end{figure}

Consider first the case of perpendicular magnetic field, $\theta=0^\circ$. From band structure we know that the extremal orbits $\alpha$ and $\beta$ are at $k_z=0$. For both orbits it is possible to form a surface $S$ which is wholly in the $k_z=0$ plane, and does not include any part of the nodal ring. Conditions (\ref{EqnTRSEnergy}) and (\ref{EqnTRSOmega}) then tell us that contributions to the surface integral cancel out pairwise, leading to zero Berry phase [Fig.~\ref{FigSM:berry_curvature}(a)]. In fact the above arguments hold even for the non-extremal orbits lying at a non-zero $k_z$.

The case of parallel magnetic field, $\theta=90^\circ$, is more complicated. Without loss of generality, align the magnetic field along the $x$-axis, so that extremal orbits lie at constant $k_x$. Because the $\gamma$ orbit does not link with the nodal ring, a surface $S$ can be formed across which the bands are always gapped [see Fig.~\ref{FigSM:berry_curvature}(b)]. Conditions (\ref{EqnMirrorEnergy}) and (\ref{EqnMirrorOmegaParallel}) mean that the flux of curvature through each face parallel to the $\hat{k}_z$-axis vanishes. For the remaining two faces whose normals point in the $\pm \hat{k}_z$ direction, conditions (\ref{EqnMirrorEnergy}) and (\ref{EqnMirrorOmegaPerpendicular}) mean that the total directed flux through these vanish as well.

On the other hand the $\delta$ orbit, which lies at $k_x=0$, links nontrivially with the nodal ring. An appropriate surface $S$ that does not contain a gapless subspace cannot be found. However Ref.~\cite{Chan2016} has formulated a relation [their Eqn.~(2.11)] between the mirror invariant and the Berry phase along the entire $k_z$ direction for a given $\bm{k}_\parallel$. Note that this phase is defined along a non-contractible loop. The key result is that this Berry phase differs by $\pi$ depending on whether $\bm{k}_\parallel$ lies inside or outside the nodal ring. Therefore we can define a contractible rectangular contour $C_R$ at $k_x=0$ with $\phi_B=\pi$, which can be deformed to the $\delta$-orbit without crossing the nodal ring of CaAgAs [Fig.~\ref{FigSM:berry_curvature}(c)]. Because of conditions (\ref{EqnMirrorEnergy}) and (\ref{EqnMirrorOmegaParallel}), the Berry phase of the $\delta$-orbit is the same as that of $C_R$.

These conclusions are summarized in Table~I in the main text. We are not aware of any strict constraints that D\textsubscript{3h}+$\mathcal{\hat{T}}$ imposes on the Berry phases of extremal orbits for generic angles. Therefore, away from these special angles, the Berry phases will in general pick up non-generic contributions. For angles close to $0^\circ$ and $90^\circ$ though, we expect these deviations to be small.


\section{The Lifshitz-Kosevich equation}

\textit{Lifshitz-Kosevich (LK) Equation.---} Under a strong external magnetic field, electronic semi-classical orbits in $k$-space are confined to planes
perpendicular to the magnetic field axis $\hat{B}$, and enclose $k$-space area $A_k$. Owing to Landau quantization they are further constrained to lie on Landau tubes, which expand as the magnetic field is increased.
As Landau tubes cross the Fermi surface, the density of states diverges, leading to oscillations in experimental quantities such as magnetic torque and magnetoresistivity. For
a three-dimensional metal $A_{k}$ will vary as a function of $k_\parallel$, where $k_\parallel$ is the component of $k$ parallel to the magnetic field.
Hence each slice at a given $k_\parallel$ will give rise to oscillations
of varying frequency and phase. The total signal will be the
sum of all these contributions. Usually the sum is
dominated by the extremal cross-sections of the Fermi surface
where $dA_{k,i} /dk_\parallel = 0$.
Therefore the fundamental frequency $F_i$ of quantum oscillations is determined by orbits on the
FS that enclose locally-extremal momentum-space area $A_{k,i}$, where $F_i=\frac{\hbar}{2\pi e}A_{k,i}$~\cite{shoenberg_1984}. Complex Fermi surfaces in general have many extremal orbits contributing to the total oscillatory response.

The grand thermodynamic potential of such a system with chemical potential $\mu$ is given by the Lifshitz-Kosevich equation \cite{shoenberg_1984}
\begin{equation}
\begin{aligned}
{\Omega}_{\rm osc} = \frac{e^{5/2} V B^{5/2}}{2^{3/2}m_e \hbar^{1/2} \pi^{7/2}}
\sum_{\text{orbits }i} \
\left| \frac{\partial^2 A_{k,i}}{\partial k_{\parallel}^2} \right|^{-1/2} \\
\sum_{p=1}^\infty p^{-5/2} R_T R_D R_S
\cdot \mathrm{cos}\left( 2\pi p \left( \frac{F_i}{B}-\frac{1}{2} + \frac{\phi_B}{2 \pi} \right)  \pm \frac{\pi}{4}  \right).
\label{EqnLifshitzKosevich}
\end{aligned}
\end{equation}
The first sum extends over all extremal Fermi surface areas $A_{k,i}$ perpendicular to the applied field, while the second sum runs over the harmonics $p$ of each fundamental oscillation frequency. The $R_T$, $R_D$ and $R_S$ are various damping terms, and the $\pm\pi/4$ is a correction that arises in 3D systems for hole carriers on an orbit with maximum or minimum cross-section respectively \cite{Li2018}.

\textit{Magnetization and Torque.---}
The magnetization parallel to the field is given by ${M}_{\parallel,\rm{osc}}=-(\frac{d {\Omega_{\rm osc}}}{dB})_{\mu}$	
\begin{equation}
\begin{aligned}
\hspace*{-3mm}
{M}_{\parallel,\rm{osc}} = -\frac{e^{5/2}  V B^{1/2}}{2^{1/2}m_e \hbar^{1/2} \pi^{5/2}}
\sum_{\text{orbits }i} \
F_i\left| \frac{\partial^2 A_{k,i}}{\partial k_{\parallel}^2} \right|^{-1/2} \\
\sum_{p=1}^\infty p^{-3/2} R_T R_D R_S
\cdot \mathrm{sin}\left[ 2\pi p \left (\frac{F_i}{B}-\frac{1}{2} + \frac{\phi_B}{2 \pi} \right)  \pm \frac{\pi}{4}  \right].
\label{EqnParallelMagnetization}
\end{aligned}
\end{equation}
The perpendicular magnetic moment from each oscillation is given by
\begin{equation}
{M}_{\perp,\rm{osc},i}= - \frac{1}{F_i} \frac{dF_i}{d \theta} \cdot M_{\parallel,\rm{osc},i}
\end{equation}	
Torque is related to the perpendicular magnetic moment and is given by
\begin{equation}
\tau = {M}_{\perp} \times B
\end{equation}	
For a three dimensional system the oscillatory torque is therefore 	
\begin{equation}
\begin{aligned}
\hspace*{-3mm}
{\tau}_{\rm osc} =\frac{e^{5/2}  V B^{3/2}}{2^{1/2}m_e \hbar^{1/2} \pi^{5/2}}
\sum_{\text{orbits }i} \
\frac{dF_i}{d \theta}\left| \frac{\partial^2 A_{k,i}}{\partial k_{\parallel}^2} \right|^{-1/2} \\
\sum_{p=1}^\infty p^{-3/2} R_T R_D R_S
\cdot \mathrm{sin}\left[ 2\pi p \left (\frac{F_i}{B}-\frac{1}{2} + \frac{\phi_B}{2 \pi} \right)  \pm \frac{\pi}{4}  \right].
\label{EqnTorque}
\end{aligned}
\end{equation}

\begin{figure}[htbp]
	\centering
	\includegraphics[trim={0cm 0cm 4.5cm 0cm}, width=0.9\linewidth,clip=true]{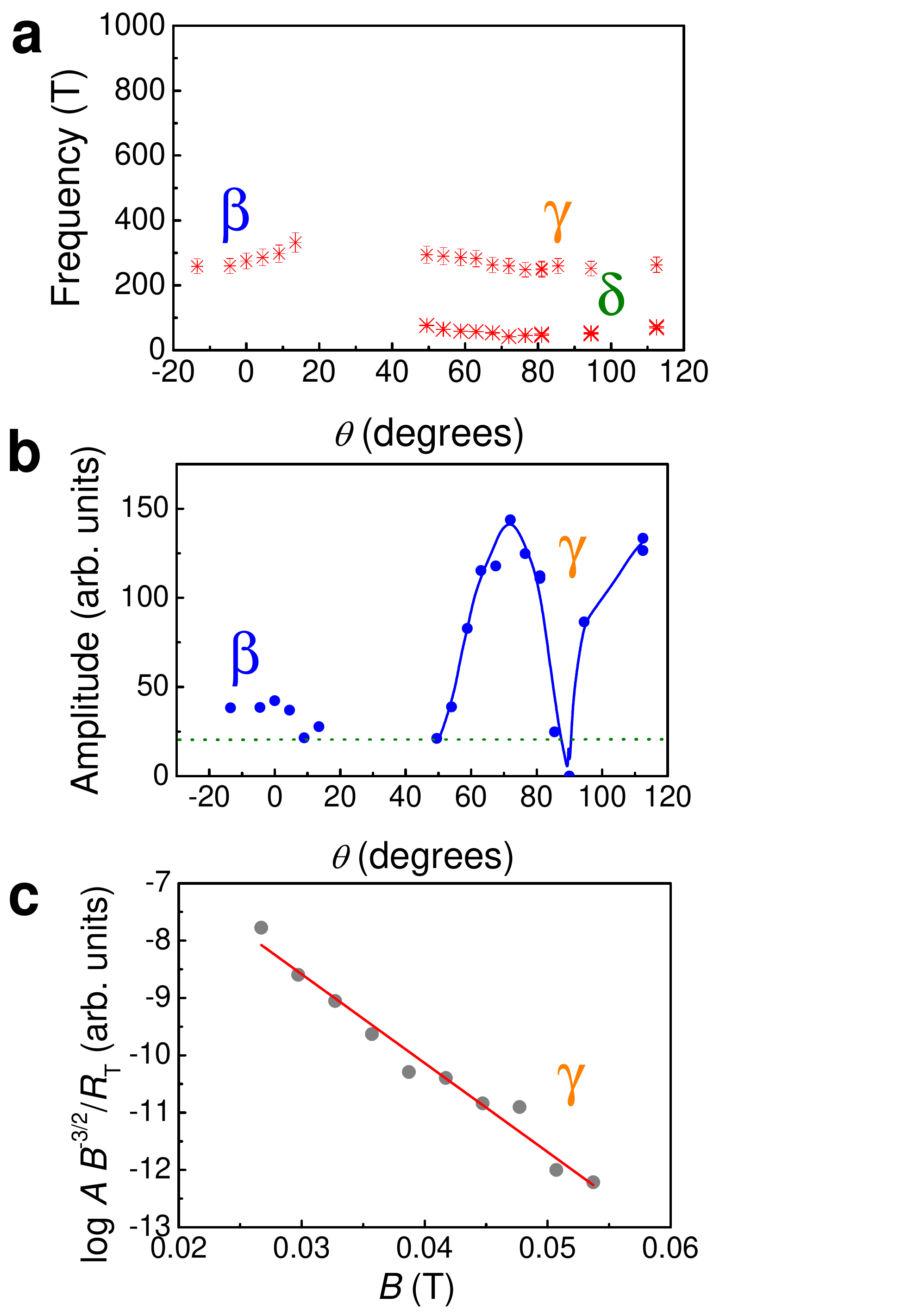}
	\caption{\textbf{Quantum oscillations parameters extracted from torque experiments.}
		The angular dependence of (a) frequencies and (b) the corresponding amplitude.
		The horizontal dotted line in (b) reflects the background noise.
		The lowest FFT peak can be influenced by other effects, such as
		the polynomial background subtraction and the $1/f$ noise and this is why direct fitting
		of data provides the best results. 
		c) The Dingle plot for a run at 0.6~K used to estimate the Dingle temperature, $T_{\rm D} \sim 65$~K. }
	\label{FigSM:QOs_experiments}
\end{figure}

\textit{Curvature Factor.---}
$C=| \partial^2 A_{k,i} /\partial k_{\parallel}^2 |^{-1/2}$
accounts for the curvature of the Fermi surface.
An extremal orbit around which the Fermi surface area hardly changes
($C$ is small), like a two-dimensional Fermi surface, will have many more states
contributing to its amplitude than an extremal orbit where the area changes sharply ($C$ is large).
The angle-dependent amplitude of quantum oscillations contained in the LK formula
depends significantly on the Fermi surface geometry, as shown in Fig.~\ref{FigSM:BS_simulations2}.
The \emph{curvature factor}
affects all types of quantum oscillations as it
reflects the density of electron orbits which contribute to the divergence of the density of states when Landau tubes cross the FS.
Furthermore, a second factor which only affects torque measurements
is the \emph{frequency factor} $ \frac{\partial F}{\partial\theta}$.
The curvature factor extinguishes the signal arising from orbits $\beta$ and $\delta$ near $\theta_c$
as shown in Fig.~\ref{FigSM:torus_simulations},
while the frequency factor for all orbits vanishes for angles close to $0^\circ$ and $90^\circ$,
as shown in Fig.~\ref{FigSM:QOs_experiments}.

\begin{figure*}[htbp]
	\centering
	\includegraphics[trim={0cm 4cm 0cm 0cm}, width=1\linewidth,clip=true]{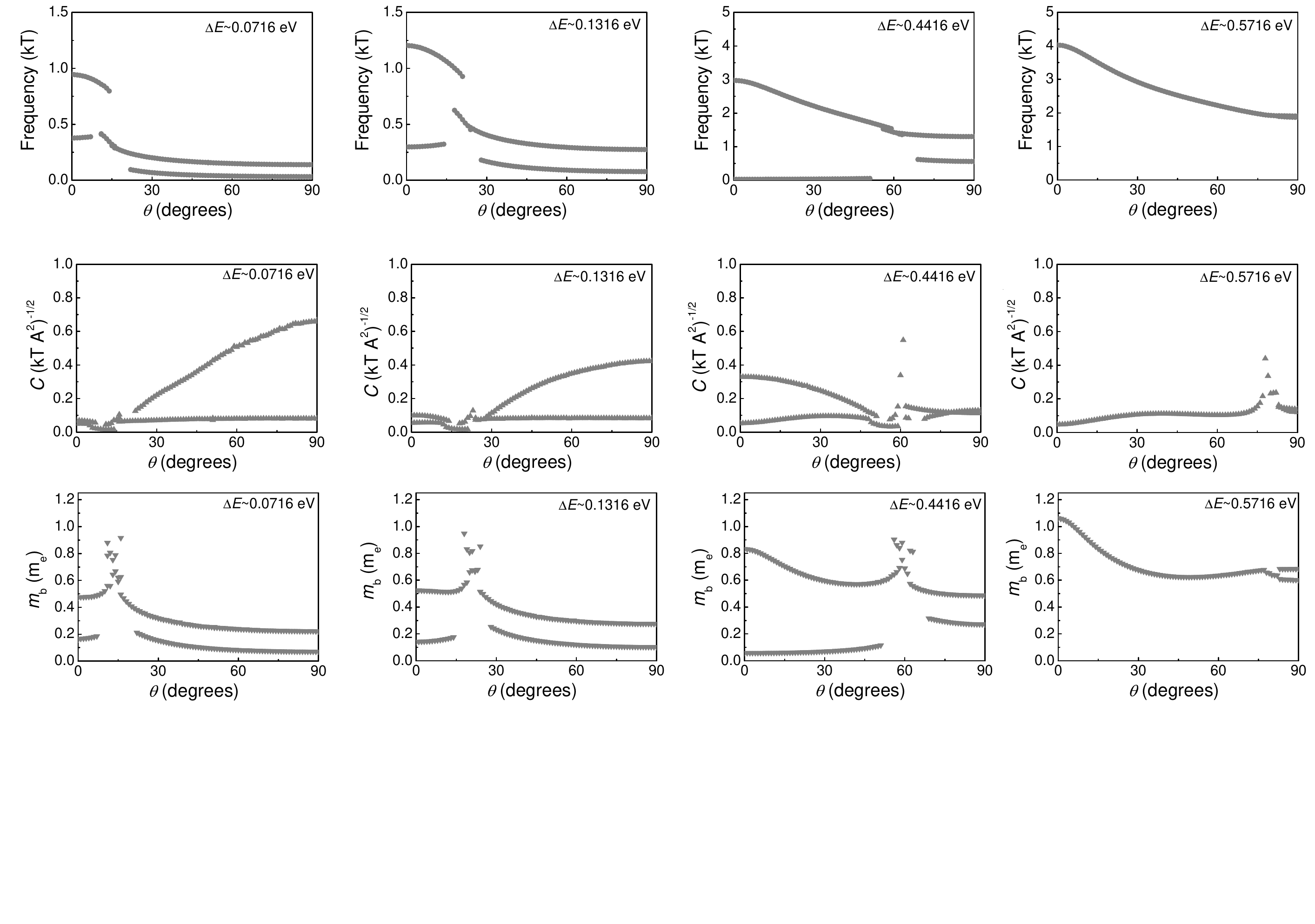}
	\caption{\textbf{Quantum oscillations parameters extracted from band structure calculations at different energies.}
		The angular dependence of simulated frequencies
		(top panel), curvature factors (middle panel) and estimated band masses (bottom panel)
		for different positions of the Fermi level of CaAgAs.	 As the system becomes more heavily doped and $\Delta E$ increases
		the torus Fermi surface become more three-dimensional and the curvature factors decreases whereas
		the calculated band masses increase.
		The orbits with the largest curvature factors and lightest masses are likely to have the largest
		amplitude in quantum oscillations.}
	\label{FigSM:BS_simulations}
\end{figure*}

\textit{Damping Terms.---} The first damping term, $R_T$, arises from the thermal broadening of the Fermi-Dirac distribution with temperature. At non-zero temperatures the step function at the Fermi level will be broadened over a range $k_B T$, and thus the Fermi surface is no longer a well-defined demarcation between occupied and unoccupied states. This {\it smearing} of the Fermi surface results in a superposition of multiple oscillations around the extremal orbit frequencies, interfering to reduce the amplitude of oscillation. This thermal damping is given by
\begin{equation}
\begin{aligned}
R_T &=\frac{X}{\mathrm{sinh}(X)} \\
\vspace{0.5cm}
X &= \frac{2 \pi^2 k_B T \, p m^*}{e \hbar B}
\end{aligned}
\label{EqnThermalDamping}
\end{equation}
where $m^*$ is the quasiparticle effective mass that is renormalized by electron-electron and electron-phonon interactions. $R_T$ depends on the ratio
$X \propto k_B T / \hbar \omega_c $, the cyclotron frequency being $\omega_c = e B/m^*$,
so the damping will be large when the thermal broadening is comparable to or greater than the spacing between adjacent Landau levels. Therefore, both low temperatures and high fields are required to reduce the thermal damping such that oscillations may be observed.
$R_T$ contains the entire temperature dependence of Equation \ref{EqnTorque}, and by fitting the temperature dependence of the amplitude, for a fixed value of field, to Equation \ref{EqnThermalDamping}, the effective mass $m^*$ can be extracted.

The Dingle term $R_D$ arises from impurity scattering of electrons
\begin{equation}
R_D = \exp\left(-\frac{\pi pm_b}{eB\tau}\right)=\exp\left(-\frac{2\pi^2pm_bk_BT_D}{e\hbar B}\right),
\end{equation}
where $\tau$ is the scattering time, $m_b$ is the band (cyclotron) mass, and $T_D=\hbar/2\pi k_B\tau$ is the Dingle temperature. Intuitively, a finite scattering time broadens otherwise sharp quantum levels into finite-width Lorentzians. Similarly to the effects of finite temperature, this smearing of the energy levels leads to a suppression of the oscillation amplitude.

\begin{figure*}[htbp]
	\centering
	\includegraphics[trim={0cm 10cm 0cm 0cm}, width=0.7\linewidth,clip=true]{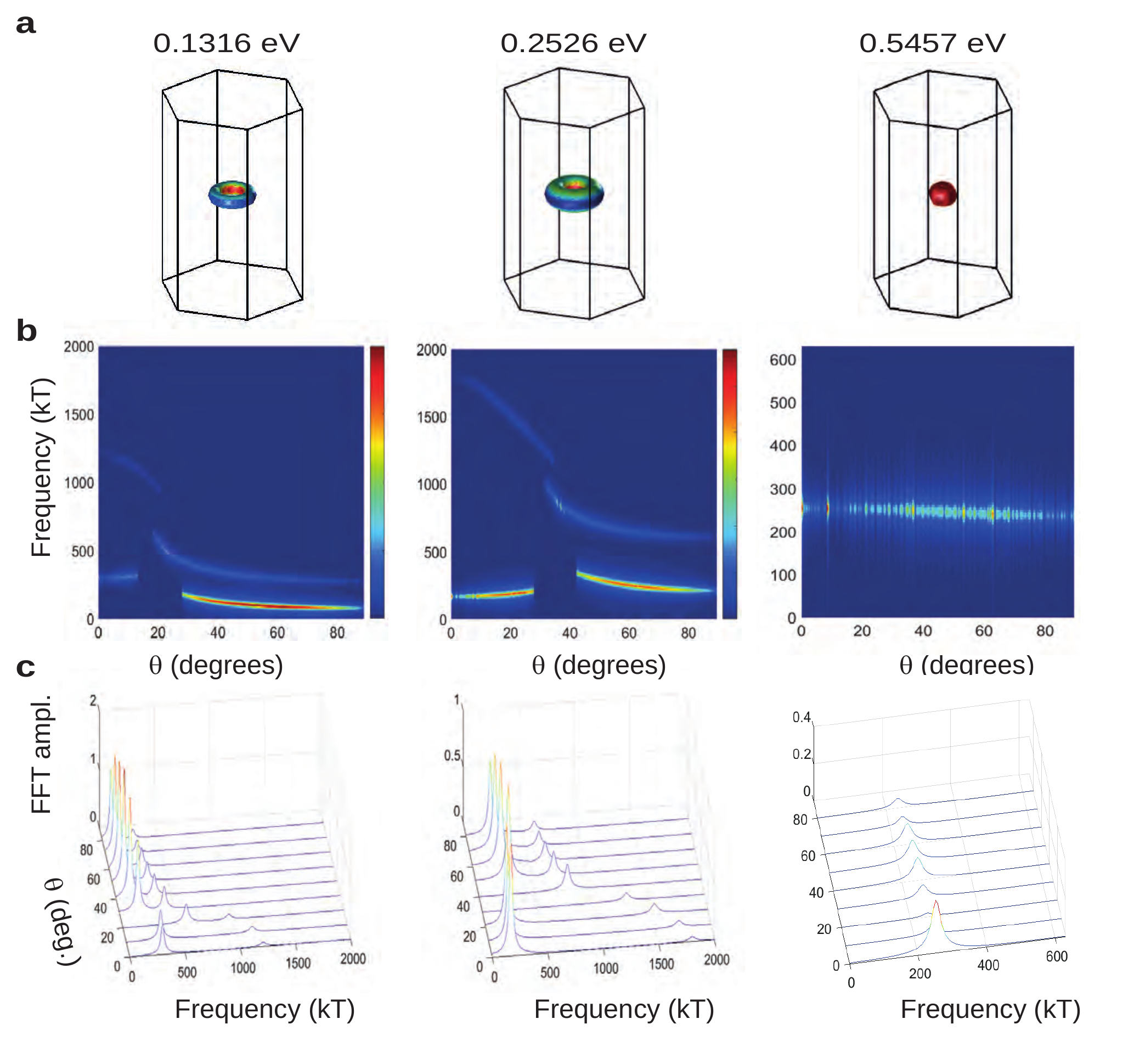}
	\caption{\textbf{Simulated quantum oscillations from band structure calculations at different energies.}
		(a) The Fermi surfaces for different shifted energies chosen to generate orbits around 260~T.
		(b) The angular dependence of simulated frequencies accounting for different
		damping terms as a colour plot or as a three-dimensional representation in (c).
		As the system becomes more heavily doped and $\Delta E$ increases
		the calculated band masses increase.
		The orbits with the largest curvature factors and lighter masses are likely to have the largest
		amplitude in quantum oscillations.}
	\label{FigSM:BS_simulations2}
\end{figure*}

Finally, the spin splitting factor $R_S$ accounts for the Zeeman splitting of Landau levels. The energy difference between the formerly spin-degenerate levels is $\Delta E=g\mu_B B$ where $g$ is the g-factor ($\simeq2$ for free electrons) and $\mu_B$ is the Bohr magneton. For a parabolic band, this is equivalent to a phase difference $2\pi\Delta E /(\hbar\omega_c)$ between the two oscillations, leading to an amplitude reduction factor due to interference
\begin{equation}
R_S = \cos\left(\frac{p\pi gm_s}{2m_e}\right),
\end{equation}
where $m_s$ accounts normally for renormalization by electron-electron interactions.
Here we assume that the electronic correlations do not play 
an important role and  $m_s \sim m^*$. 
This assumes that the pair of spin-split oscillations have equal amplitudes and decay envelopes. If the phase difference for $p=1$ is an odd multiple of $\pi$, a spin zero manifests as an extinction of the dominant harmonic. This can detected in materials with a large and anisotropic g-factor~\cite{Wang2018}. This will also affect Berry phase estimates, because a negative $R_S$ for $p=1$ appears as a phase inversion in the dominant oscillation.

Assuming that the 260~T frequency peak and its small 210~T shoulder in Figure~2(b) (main text)
arises from spin-splitting, the $g$-factor of CaAgAs at that orientation can be estimated. 
According to band structure calculations, these frequencies can be obtained by
hole doping levels of $0.132$~eV and $0.115$~eV respectively, i.e.~a Zeeman splitting of $\Delta E=0.017$~eV. 
Taking a representative field strength of $B=40$~T, this leads to $g=7.3$ at $\theta=74^\circ$. 
One would expect  a spin zero effect when $R_S$ goes to zero and $g m^*/m_e$ is close to 1.
The amplitude of $\gamma$ orbit drops close to $\theta \sim 50^{\circ}$ in 
Fig.~\ref{FigSM:QOs_experiments}(b) where a spin zero effect may occur.
The suppressed amplitude at $90^\circ$ is due to the frequency factor $dF/d\theta$ vanishing at high-symmetry angles.

%

\section{Circular Torus Model}

In this section we outline the circular torus model, which is an idealization of the FS of CaAgAs at finite doping. The model is the surface of revolution about the $k_z$-axis of a circle of radius $r$ centered at a distance $R$ from the axis, and satisfies
\begin{equation}\label{EqnTorusModel}
(\sqrt{k_x^2+k_y^2}-R)^2+k_z^2=r^2.
\end{equation}
The major and minor radii $R$ and $r$ are related to the radius of the nodal ring and the doping level respectively. The $k_z$-axis is aligned with the $c$ axis of the CaAgAs crystal.

\begin{figure}[htbp]
	\centering
	\includegraphics[trim={0cm 0cm 0cm 0cm}, width=1\linewidth,clip=true]{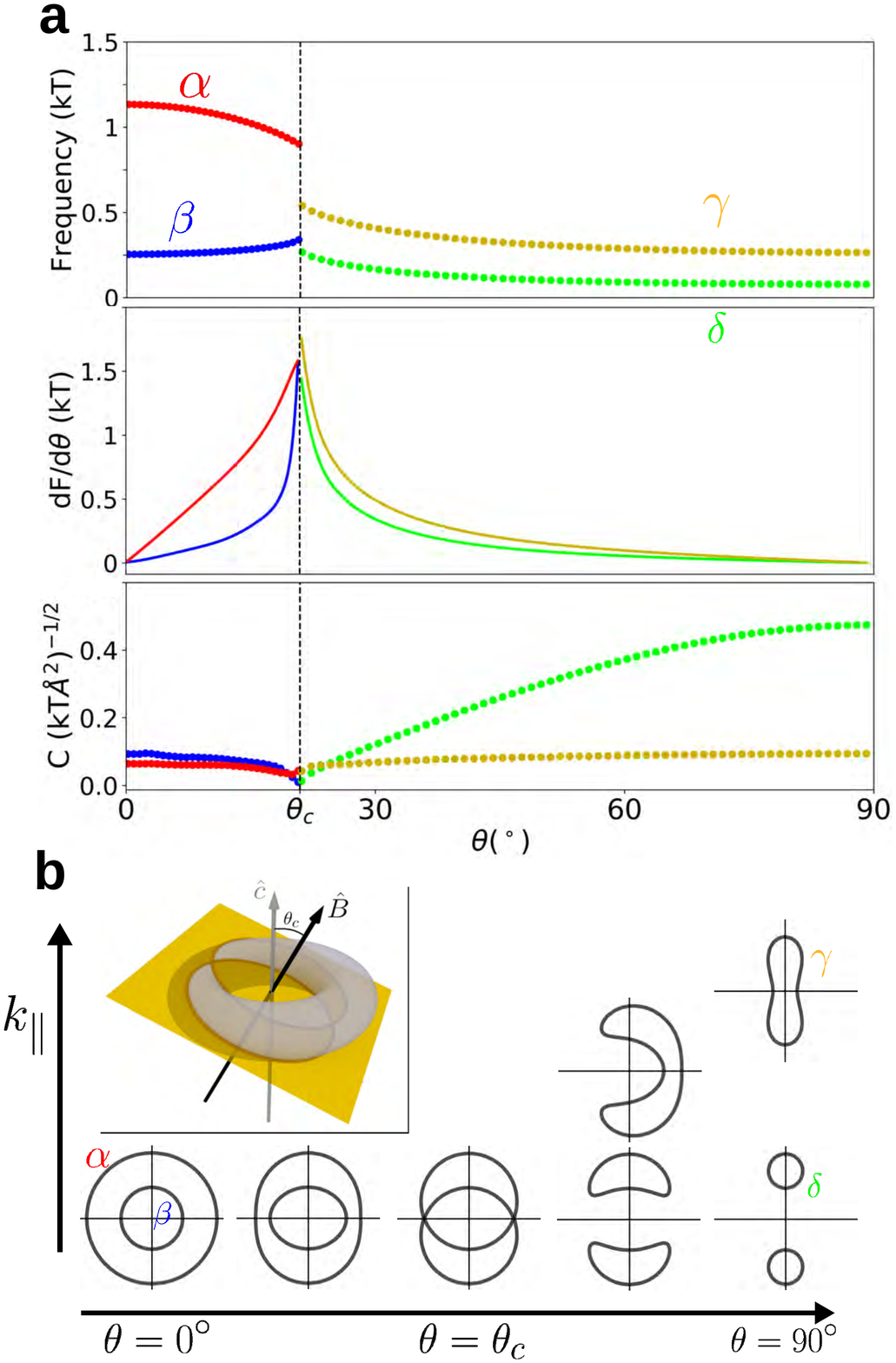}
	\caption{\textbf{Circular torus model}. (a) The top panel shows the frequencies for the four orbits, the middle panel shows the magnitude of the frequency factor, and the bottom panel shows the curvature factor. The parameters $R/r=2.8$ ($\theta_c=20.9^\circ$) are chosen so that the frequencies at $\theta=0^\circ,90^\circ$ are consistent with band structure calculations at a hole doping of 0.1316~eV. (b)Torus cross-sections that contain extremal orbits, as a function of angle $\theta$. Inset shows the critical angle condition and the resulting Villarceau circles that form at $k_\parallel=0$. }
	\label{FigSM:torus_simulations}
\end{figure}

We are interested in planar cross-sections (toric sections) whose normal is at an angle $\theta$ to the $k_z$-axis. These contain the semi-classical orbits traversed by electrons in the presence of an external magnetic field $B$. The distance of such a plane to the origin is denoted $k_\parallel$. Since there is rotational symmetry, we can choose the normal to lie in the $k_x$-$k_z$ plane. Expressing Equation \ref{EqnTorusModel} in terms of in-plane coordinates $(k_x',k_y')$, we obtain
\begin{equation}\label{EqnToricSection}
(k_x^{\prime2}+k_y^{\prime2}+k_\parallel^2-r^2)^2=4R^2\left(\left(k_x^{\prime2}\cos\theta+R\sin\theta\right)^2+k_y^{\prime2}\right).
\end{equation}
For every value of $(k_\parallel,\theta)$, Equation \ref{EqnToricSection} parameterizes a toric section. Disconnected pieces of a toric section give rise to distinct semi-classical orbits. The areas $A_k$ of these orbits can be numerically calculated for different $(k_\parallel,\theta)$, and the locally extremal orbits for a fixed $\theta$ will give rise to quantum oscillations at the magnetic field orientation described by $\theta$. The curvature factor $C=\left|\partial^2A_k/\partial k_\parallel^2\right|^{-1/2}$ and frequency factor $\partial F/\partial\theta$ can also be computed in this way. Figure \ref{FigSM:torus_simulations}(a) shows an example set of results, where the parameters have been chosen to to match band structure calculations at a hole doping of 0.1316~eV.

It is found that the circular torus has a critical angle $\sin\theta_c=r/R$. For $\theta<\theta_c$ there are two non-degenerate extremal orbits at $k_\parallel=0$. At $\theta>\theta_c$, there are two extremal orbits, one at $k_\parallel=0$ and a larger one at $k_\parallel\neq0$. Each is doubly degenerate. Near the critical angle, the larger orbit lies at a plane $k_\parallel\sim\theta-\theta_c$.  At $\theta=\theta_c$, the various orbits merge to form a set of intersecting Villarceau circles at $k_\parallel=0$. The orbits are labelled $\alpha,\beta,\gamma,\delta$ according to Figure 1(a) (main text) and Figure \ref{FigSM:torus_simulations}(b).
All the orbits have a curvature factor that decreases as the critical angle is approached [Fig.~\ref{FigSM:torus_simulations}(a)]. The $\beta$ and $\delta$ orbits have a vanishing curvature factor at $\theta_c$. All the orbits have a vanishing frequency factor for angles close to the high-symmetry orientations $0^\circ$ or $90^\circ$.

The volume of the torus is $V_k=(\pi r^2)(2\pi R)$, which allows for an estimate for the carrier density $n=2V_k/(2\pi)^3$ where the factor of 2 arises from spin.


\section{Phase analysis of quantum oscillations }

\begin{figure*}[htbp]
	\centering
	\includegraphics[trim={0cm 18cm 1cm 0cm}, width=0.8\linewidth,clip=true]{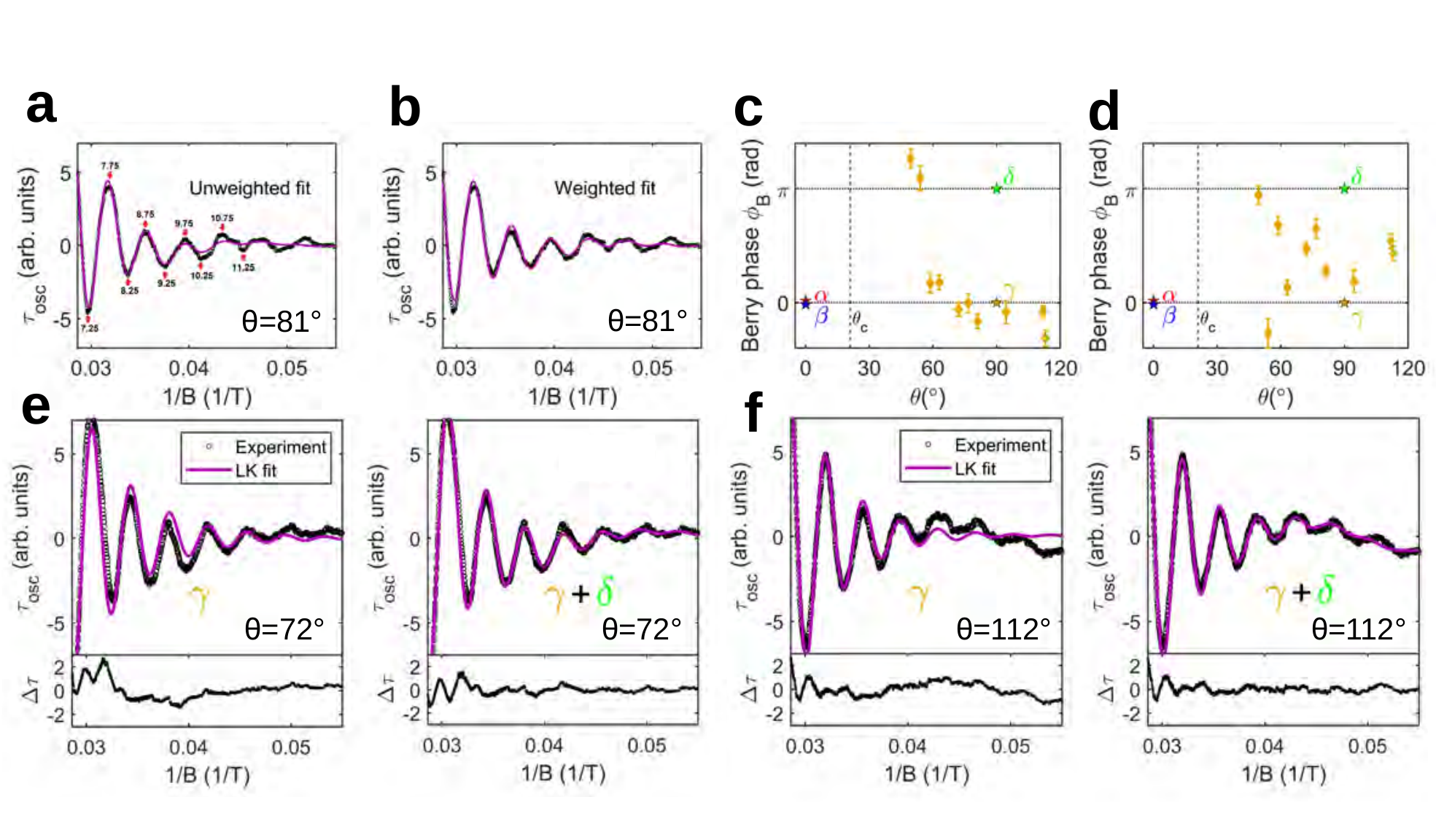}
	\caption{\textbf{Phase analysis of quantum oscillations of CaAgAs.} (a) LK fitting to oscillatory signal without weighting. Numbered red arrows indicate the identification of Landau level indicies $n$ required for the Landau fan. (b) LK fitting to oscillatory signal with larger weight on high $1/B$ data, in order to capture a larger number of oscillations. (c) Berry phases for the $\gamma$ orbit derived from the Landau fan analysis [Fig.~\ref{FigPhases}(b)] where the slopes are fixed via the LK fitting procedure. (d) Same as (c) except the slopes are fixed via the FFT peaks. (e) One-component and two-component LK fits for $\theta=72^\circ$, which predicts a Berry phase of $\phi_B=0.6\pi$ for the $\delta$ orbit. (f) An example of a dataset ($\theta=112^\circ$) where the quadratic subtraction contaminates the $\delta$ oscillation. The resulting fit gives an unphysical negative Dingle temperature, so the predictions for $\delta$ are discarded.}
	\label{FigSMdataanalysis}
\end{figure*}
\textit{Background Subtraction.---} A quadratic polynomial is used to extract the oscillatory signal from the raw torque data. The bulk torque response is expected to scale quadratically with field since $\bm{\tau}=\mathbf{M}\times\mathbf{B}$. We avoid a high-order polynomial subtraction in order to prevent over-fitting to the low frequency $\delta$ orbit.

\textit{Lifshitz-Kosevich Fitting.---} The subtracted data for large angles $\theta>50^\circ$ is modeled with a simplified two-component Lifshitz-Kosevich equation, see Equation \ref{EqnTorque}. As the operating temperature for the angle-dependent field sweeps was low ($T$=0.5~K),
the temperature factor $R_T$ is neglected in the direct fitting. The spin-splitting factor $R_S$ only affects the overall amplitude and is absorbed into the overall amplitude $A_i$ along with all other constant prefactors. Higher harmonics are not included since they do not appear to be significant in the Fourier spectrum. There is an additional overall minus sign because $dF/d\theta$ is negative for both $\gamma$ and $\delta$ orbits.

The fitting is performed using weighted and constraint-free non-linear least squares over the field range 18-35T where there are discernible oscillations. The weights are implemented because unweighted fitting tends to focus on the large amplitude oscillations at small $1/B$ at the expense of the oscillations at high $1/B$ and the error bars in Fig.~3(a) indicate 95\% confidence intervals 
[Fig.~\ref{FigSMdataanalysis}]. It is verified that the weighting does not substantially change the predicted frequencies and phases. The parameters are initialized to values estimated from the FFTs to ensure convergence. The characteristics of the $\gamma$ orbit are well characterized by this method. However the $\delta$ orbit has relatively weaker and fewer oscillations, and is therefore more prone to being masked by noise and the polynomial subtraction [Fig.~\ref{FigSMdataanalysis}(f)]. We only keep the estimates for $\delta$ if its parameters have converged to physically sensible values and the resulting fit looks reasonable.
Over the 18-35~T field range only two periods of the $\delta$ orbit could be resolved,
and their amplitude is weak compared to the $\gamma$ orbit.
This is reflected in the spread and relatively large error bars in the estimates of $\phi_B$ [Fig.~3(a)].
On the other hand, the $\gamma$ orbit has stronger oscillations and a higher number of discernible periods (see Fig.~\ref{FigSMdataanalysis}). 
This allows us to conclude that $\gamma$ carries zero Berry phase near $\theta=$~90$^\circ$. 

\textit{Landau Fan Analysis.---} The positions of the maxima and minima of quantum oscillations can be used to extract the Berry phase through a Landau fan diagram [Fig.~3(b)]. Taking into account the negative $dF/d\theta$, we label the signal peaks with $F/B=3/4,7/4,11/4,\ldots$ and valleys with $F/B=1/4,5/4,9/4,\ldots$ (this is the Landau index $n$), and plot $n$ vs $1/B$. Fitting a straight line $n=\alpha_1/B + \alpha_2$ gives the frequency as $F=\alpha_1$ and the phase as $\phi=2\pi\alpha_2$.

The $\delta$ orbit is not amenable to such an analysis because the peaks/valleys cannot be identified by eye. For the $\gamma$ orbit, we do not reach the quantum limit and the lowest Landau index is quite high $n\sim7$ [Fig.~\ref{FigSMdataanalysis}(a)]. This results in large errors for the estimate of $\phi$ due to the uncertainty in the slope. To mitigate this, we fixed the slope of the linear fits to the frequencies obtained from either the Lifshitz-Kosevich fitting procedure [Fig.~\ref{FigSMdataanalysis}(c)] or the FFT peak position [Fig.~\ref{FigSMdataanalysis}(d) and Fig.~\ref{FigSM:QOs_experiments}(a)]. The Berry phases are extracted using $\phi_B = \phi+3\pi/4$. 
The FFT spectra gives slightly lower frequencies (within errors)
then the direct LK fitting procedure, 
and hence the predicted Berry phases are larger in Fig.~\ref{FigSMdataanalysis}(d).

\begin{figure*}[htbp]
	\centering
	\includegraphics[trim={0cm 0cm 0cm 0cm}, width=0.6\linewidth,clip=true]{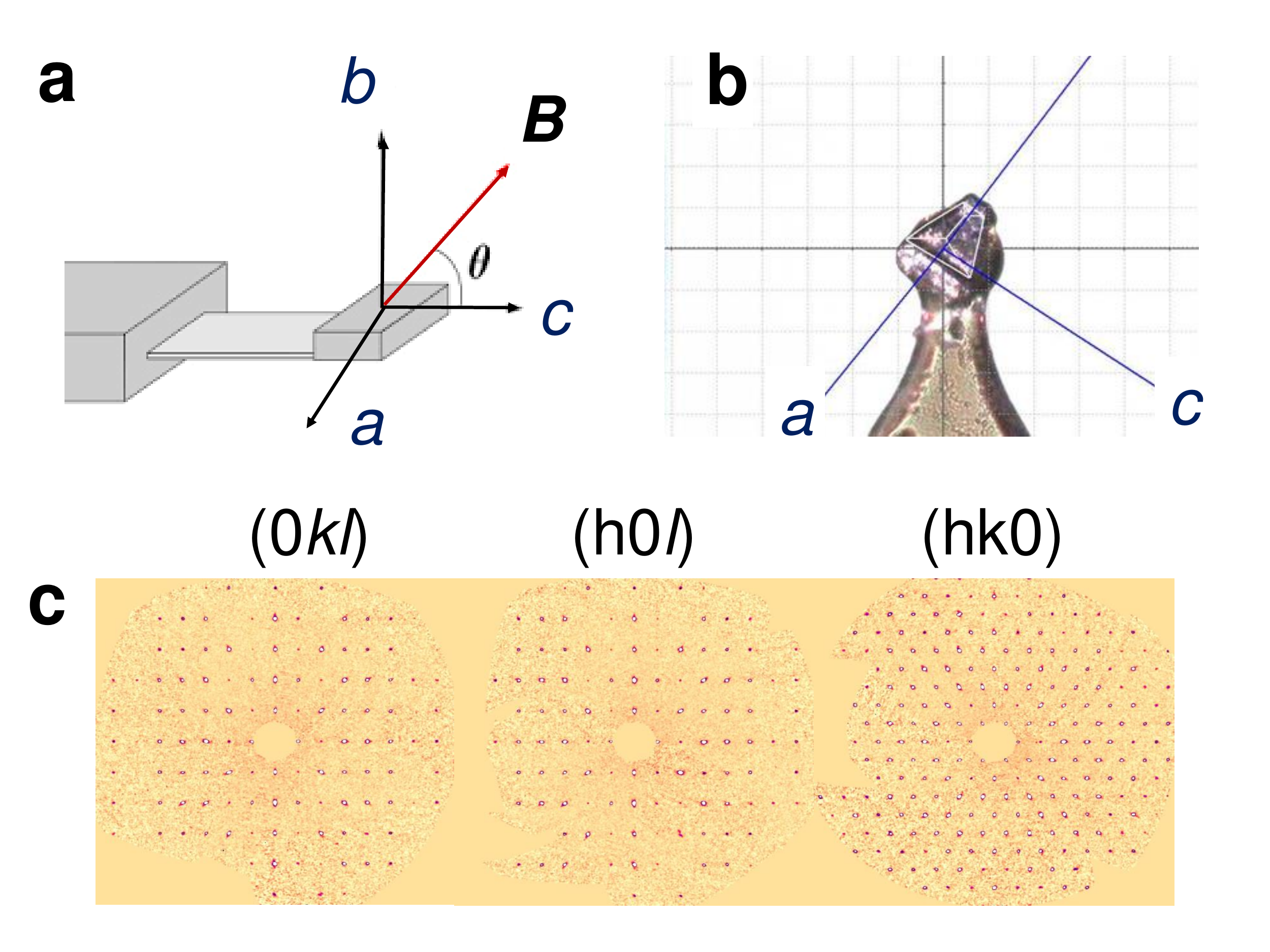}
	\caption{\textbf{Sample orientation and X-ray of a single crystal of CaAgAs used in torque measurements.}
		(a) Schematic representation of a sample on a level defining the orientation in magnetic field. (b) A picture of the crystallographic axes in relation
		to the crystal edges. (c) Laue X-ray diffraction patterns along different crystallographic planes indicating
		the single phase nature of the measured crystal. The measured lattice parameters
		are $a$ = $b$ = 7.2040(6)~\AA~ and $c$ = 4.2700(4) \AA~
		for a hexagonal symmetry group P$\bar{6}$2m (189).  }
	\label{FigSM:Xrays}
\end{figure*}

\newpage
\clearpage

\bibliography{CaAgAs_bibjun19}

\end{document}